# Enhanced Sufficient Battery Model for Aggregate Flexibility of Thermostatically Controlled Loads Considering Coupling Constraints

Guangrui Wang, *Student Member, IEEE*, Zhengshuo Li, *Member, IEEE*

*Abstract*--This letter proposes an enhanced sufficient battery model (ESBM) as well as a binary search algorithm for a sharp inner-approximation of the aggregate flexibility of thermostatically controlled load (TCL) arrays. Compared with the previous work on generalized battery models, this ESBM preserves the merits of being sufficient and mitigates the conservativity. Moreover, unlike the work ignoring the coupling constraints that may also restrict TCLs' aggregate flexibility, our ESBM can readily handle these constraints. Numerical tests validate the merits of using the ESBM and its significance for power system operations.

*Index Terms*—Aggregation, flexibility, generalized battery model, thermostatically controlled load.

## I. Introduction

NUMEROUS small-scale thermostatically controlled loads (TCLs) are often aggregated to provide balancing services for power systems. Several modeling methods for the aggregate flexibility of an array of TCLs have been proposed in the literature [1]-[5]. Among these methods, He *et al.* [5] proposed generalized battery models to approximately represent the realistic aggregate flexibility (RAF) of an array of TCL based on inner- and outer-approximation, referred to as a sufficient battery model (SBM) and a necessary battery model (NBM), respectively. With their method, the flexibility of individual TCLs can be aggregated to balance renewable-integrated (e.g., wind energy) power systems. However, as illustrated in Fig.1(a), the SBM in [5] can be conservative in the sense that its boundary might not be *sharp*: there is a positive inner distance to the boundary of RAF. Although their later work [6] improved the sharpness of this inner-approximation for each individual TCL, it should be noted that [6] neglected the underlying coupling constraints regarding the TCL array, and these constraints may restrict the aggregate flexibility. For example, a typical coupling constraint is that the aggregate flexibility of the TCLs located downstream from a line should be limited by the transmission capacity of the upstream lines. Apparently, when such a coupling constraint becomes active, the Minkowski sum of the (even sharp) inner-approximation regarding every TCL, e.g., as suggested in [6], may deviate from the RAF of the TCL array.

In this letter, we present an *enhanced sufficient battery model* (ESBM) to represent the aggregate flexibility of an array of TCLs, which also seeks for a sharp inner-approximation: the boundary of the ESBM should touch that of the RAF at least at one point. One notable feature of our ESBM is that it directly enhances the sharpness of the SBM of the *ensemble* of the TCL array rather than an *individual* TCL that was conducted in [4],[6]. Hence, the underlying coupling constraints regarding the TCL array, e.g., the preceding line flow constraint, can be readily considered and handled with our ESBM. The case study will show that our ESBM can more accurately evaluate the aggregate flexibility of TCL arrays as well as its significance for power system operations.

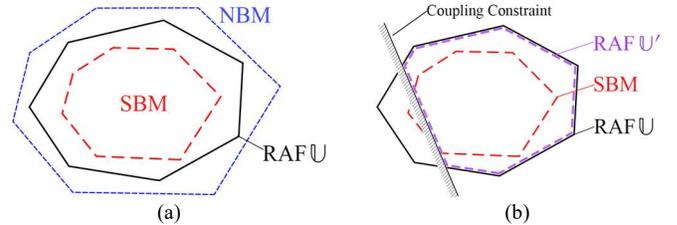

Fig. 1. Illustration of the basic concepts: (a) SBM, NBM, and RAF; (b) how an active coupling constraint affects the RAF.

## II. Enhanced Sufficient Battery Model

### A. Preliminaries: SBM and NBM

As per [5], an individual TCL $i$'s power $p_t^i$ is divided into two parts: one is the baseline power $p_b^i$ to keep the TCL working at its set-point temperature $\theta_r^i$, and the other, $u_t^i = p_t^i - p_b^i$, represents the flexibility that the TCL can provide to the grid. The temperature evolution [6] of this TCL is

$$x_t^i = \gamma x_{t-1}^i + \delta u_t^i , \qquad (1)$$

where $x_t^i = C_{th}^i(\theta_r^i - \theta_t^i)/\eta$ can be regarded as the thermal storage of TCL $i$; $\gamma = e^{-\Delta T/R_{th}^i C_{th}^i}$ and $\delta = (1-\gamma)R_{th}^i C_{th}^i$ are coefficients; $\theta_t^i$ is the temperature of this TCL at time $t$; $C_{th}^i$, $R_{th}^i$, and $\eta$ are the thermal capacitance, thermal resistance, and the coefficient of performance, respectively; $\Delta T$ is the schedule time interval.

The summation of $u_t^i$ for every TCL builds up the aggregate flexibility of a TCL array. In [5], He *et al.* proposed the concept of a generalized battery model $B(\phi)$ to model the RAF of an array of TCLs, denoted by $\mathbb{U}$. The charge or discharge rate of this generalized battery represents the aggregate power that the TCL array exchanges with the grid. This generalized battery model $B(\phi)$ is formally defined by parameters $\phi = (C, n_+, n_-, \alpha)$, where $C$ is the battery energy capacity, $n_+$ and $n_-$ are discharge and charge rate limits, respectively,

This work was supported by National Natural Science Foundation of China under Grant 52007105. Guangrui Wang and Zhengshuo Li are with the School of Electrical Engineering, Shandong University, Jinan, China. Zhengshuo Li is the corresponding author (email: zsli@sdu.edu.cn).

and $\alpha$ is the dissipation rate. Next, in the absence of the coupling constraints, He *et al.* provided two generalized battery models to bound $\mathbb{U}$ as follows:

$$B(\phi_1) \subseteq \mathbb{U} \subseteq B(\phi_2). \qquad (2)$$

The inner-approximation model $B(\phi_1)$ is called an SBM and the outer $B(\phi_2)$ called an NBM, because every point $r \in B(\phi_1)$ is feasible, i.e., inside $\mathbb{U}$, while $B(\phi_2)$ may contain infeasible points that are outside $\mathbb{U}$, as illustrated in Fig.1(a).

He *et al.* then showed that with different criteria of offering flexibility, there are several types of SBMs and NBMs. In this letter, we take as an example the SBM and NBM based on the criterion of offering the largest charge rate limit $n_-$, whose parameters are computed as per [5] and listed in Table I. Nevertheless, it will be easy to adapt our ESBM to the other types of SBMs and NBMs with the other criteria like offering the largest discharge rate limit $n_+$ or battery energy capacity $C$.

TABLE I. FORMULAE OF PARAMETERS OF GENERALIZED BATTERY MODELS

|       | $B(\phi_1)$ | $B(\phi_2)$ |
|-------|-------------|-------------|
| $n_+$ | $n_{1,+} = (\sum_i p_b^i) \min_i \frac{p_m^i - p_b^i}{p_b^i}$ | $n_{2,+} = \sum_i (p_m^i - p_b^i)$ |
| $n_-$ | $n_{1,-} = \sum_i p_b^i$ | $n_{2,-} = \sum_i p_b^i$ |
| $C$   | $C_1 = (\sum_i p_b^i) \min_i \frac{f^i}{p_0^i}$ | $C_2 = \sum_i (1 + |1 - \frac{a^i}{\alpha}|) \frac{\Delta \theta^i}{b^i}$ |

In Table I, $p_m^i$ is the rated power of TCL $i$; $a^i$ and $b^i$ are two factors related to $C_{th}^i$, $R_{th}^i$, $\alpha$ and $\eta$; $f^i$ is a parameter defined by $a^i$, $b^i$, $\alpha$, and the temperature dead-band $\Delta\theta^i$. Detailed deviations of the formulae are referred to [5].

*B. General Idea of ESBM*

As shown in Fig.1(a), the SBM, an inner-approximation of $\mathbb{U}$, may not be sharp. On the one hand, one can observe that $\mathbb{U}$ lies between the SBM and NBM when the coupling constraints are inactive. Hence, if we define a new model $B(\phi_3)$ that is a convex combination of the SBM and NBM, namely $B(\phi_3) = (1-\mu)B(\phi_1) + \mu B(\phi_2)$ where $\mu \in [0,1]$, then there should be a $\mu$ such that $B(\phi_3)$ is a sharp inner-approximation of $\mathbb{U}$. Since $B(\phi_3) = B(\phi_1)$ when $\mu = 0$, this constructed $B(\phi_3)$ can be deemed an enhanced SBM that both preserves the merits of *being sufficient* and mitigates possible conservativity, so we call this $B(\phi_3)$ an ESBM. On the other hand, notice that if certain coupling constraints become active, the RAF may shrink from $\mathbb{U}$ to a certain $\mathbb{U}'$ such that $B(\phi_1)$ is no longer sufficient, as is illustrated in Fig.1(b). In this case, the ESBM should be formulated as $B(\phi_3) = (1+\mu)B(\phi_1)$ where $\mu \in [-1,0]$ instead, but there should also be a $\mu$ such that this ESBM shrinks from the SBM $B(\phi_1)$ to a sharp inner-approximation of $\mathbb{U}'$. Lastly, note that these two formulations of $B(\phi_3)$ are consistent when $\mu = 0$.

*C. Formulation of ESBM and Algorithm*

Following the above idea, it is easy to see that computing the parameters $\phi_3$ of the ESBM depends on the sign of $\mu$. For example, the discharge rate limit $n_+$ of the ESBM is formulated as $(1-\mu)n_{1,+} + \mu n_{2,+}$ when $\mu \in [0,1]$, and as $(1+\mu)n_{1,+}$ when $\mu \in [-1,0]$, where $n_{1,+}$ and $n_{2,+}$ are the counterparts of the SBM and NBM, respectively. The computation of the other parameters in $\phi_3$ follows a similar rule. Since one cannot *a priori* know the sign of $\mu$, one needs to first check whether $B(\phi_3)$ associated with $\mu = 0$, namely $B(\phi_1)$, is sufficient. If the answer is yes, then choose the formulation of ESBM associated with $\mu \in [0,1]$, and choose the other formulation otherwise.

Second, for either formulation of the ESBM, compute the maximum $\mu$ such that the associated $B(\phi_3)$ is sufficient. This formally requires that every point within the boundary of $B(\phi_3)$ associated with a fixed $\mu$ should lie inside the RAF. Physically, it means that the system operator's every possible instruction on the aggregate power of the TCL array should be feasibly disaggregated among the individual TCLs without violating any TCL's operating limit and coupling constraint.

This problem can mathematically be formulated as: $\max \mu$ such that $Q(\mu) = 0$, where $Q(\mu)$ is the measure of the maximal outer distance to the RAF given a fixed $\mu$, defined as

$$Q(\mu) = \underset{P_{sys}}{\text{maximize}} \ \underset{u^i, x^i, l^+, l^-}{\text{minimize}} \ l^+ + l^-, \qquad (3.1)$$

where the feasible region of the inner-problem is constrained by the temperature evolution (1) for every TCL and

$$P_{sys} = \sum_i u^i + l^+ - l^-, \qquad (3.2)$$

$$l^+, l^- \geq 0, \qquad (3.3)$$

$$-p_b^i \leq u^i \leq p_m^i - p_b^i, \qquad (3.4)$$

$$-\Delta\theta^i C_{th}/\eta \leq x^i \leq \Delta\theta^i C_{th}/\eta, \qquad (3.5)$$

$$f_{min}^l \leq \mathbf{H}^l(\boldsymbol{p_b} + \boldsymbol{u}) \leq f_{max}^l, \forall l \in N^l. \qquad (3.6)$$

As for the outer problem in (3.1), the system operator's instruction $P_{sys}$ should be subject to the admissible set of the ESBM, namely, $P_{sys} \in (1-\mu)B(\phi_1) + \mu B(\phi_2)$ when $\mu \in [0,1]$, and $P_{sys} \in (1+\mu)B(\phi_1)$ when $\mu \in [-1,0]$.

Notice that the subscript $t$ of $u_t^i$ and $x_t^i$ in (1) is now omitted in (3) for simplicity; $\boldsymbol{p_b}$ and $\boldsymbol{u}$ are the vectors of $p_b^i$ and $u^i$; $l^+$ and $l^-$ are slack variables representing how far this $P_{sys}$ deviates from the RAF; $\mathbf{H}^l$ is the distribution factors of the power through line $l$ with regard to the output of the TCLs; $f_{max}^l$ and $f_{min}^l$ are the forward and reverse flow limit on the TCLs' power for line $l$; $N^l$ is the set of the lines.

Equation (3.2) indicates that the system operator's power instruction on the ensemble of the TCLs should be tracked by the aggregate output of the individual TCLs with the minimal mismatches (being zero if possible) denoted by $l^+ - l^-$, as implied by the inner minimization in (3.1). Constraint (3.3) enforces the nonnegativity of the slack variables. Constraints (1), (3.4), and (3.5) show the operating limits for every individual TCL [6]. Constraint (3.6) imposes limits on the aggregate output of the TCLs located downstream from line $l$, whose derivation can be referred to the appendix in [7]. Notice that many other coupling constraints can also be represented by (3.6) if they can be modeled via linear functions of $\boldsymbol{u}$.

Directly solving the maximal $\mu$ subject to $Q(\mu) = 0$ is mathematically challenging. However, given a fixed $\mu$, (3) can be readily solved with the method in [8] after taking the dual over the inner minimization. Therefore, one can exploit a binary search algorithm to seek a suboptimal solution of $\mu$: if the stopping criterion is set to $10^{-2}$, only eight rounds of bisection and solving (3) are required. One may also be interested in its parallel variant as shown in Table II. Here, given an appropriate computing environment, one can evaluate all the candidate $\mu_k \in [-1,1]$ in parallel prior to an expeditious binary search over these evaluated $Q(\mu_k)$, which will further reduce the overall computational time. A mathematical sounder algorithm to solve the optimal $\mu$ will be studied in our future work.

TABLE II. A PARALLEL VARIANT OF BINARY SEARCH ALGORITHM

**Algorithm**: Binary search for a suboptimal $\mu$ (interval set to, e.g., 0.01).
1: Calculate the parameters of $B(\phi_1)$ and $B(\phi_2)$ as per Table I.
2: Let $K = 2/\text{interval}$. Define a sequence of number $\{\mu_k\} \in [-1,1]$, $k = 0,1,\ldots,K$ such that $\mu_{k+1} - \mu_k = \text{interval}$.
3: Solve (3) for all $\mu_k \in \{\mu_k\}$ in parallel, and generate the sequence of $\{Q(\mu_k)\}$.
4: **repeat** (start with $a = 0$ and $b = K$):
5: $c = [(a+b)/2]$, where $[x]$ is the least integer function.
6: **if** $Q(\mu_a) = 0$ and $Q(\mu_c) > 0$, then let $b = c$.
7: **if** $Q(\mu_c) = 0$ and $Q(\mu_b) > 0$, then let $a = c$.
8: **until** $b - a = \text{interval}$, then let $\mu^* = \mu_a$.
9: **output**: if $\mu^* \in [-1,0]$, $B(\phi_3) = (1+\mu^*)B(\phi_1)$ or
   if $\mu^* \in [0,1]$, $B(\phi_3) = (1-\mu^*)B(\phi_1) + \mu^*B(\phi_2)$.

## III. CASE STUDY

An IEEE 14-bus system with three thermal generators and one wind farm was simulated in a robust rolling economic dispatch program [8] for 24 hours with the schedule time interval $\Delta T$ being 10 minutes. The daily load demand varies from 189 MW to 339 MW, with an average of 258 MW. The output of the wind farm varies in the range of 0 to 75 MW. There are four TCL arrays on bus 6, each having 1000 to 4000 TCLs. These TCL arrays are aggregated together to provide flexibility. The computed parameters of NBM, SBM, and ESBM are listed in Table III. Here, let ESBM1 denote the result when there is no active coupling constraint; ESBM2 the result with active line flow constraints; ESBM3 the result with tighter line flow limits. These three ESBMs are compared to demonstrate the impact of the coupling constraints on the aggregate flexibility of TCL arrays. In addition, the last column of Table III, entitled $\eta_w$, represents the percentage utilization of the wind power in the outcome of the dispatch program. This is to show to what degree the different modeling of the flexibility of TCL arrays will affect the outcome of the power system operations and, in turn, to indicate the significance of leveraging the ESBM.

TABLE III. THE RESULTS WITH DIFFERENT SIMULATION SETTINGS

| Model | $\mu$ | $n_+$ | $n_-$ | $C$ | $\eta_w$ |
|---|---|---|---|---|---|
| NBM | 1.00 | 30.31 MW | 15.29 MW | 6.28 MW·h | 67.8% |
| SBM | 0.00 | 19.12 MW | 15.29 MW | 2.52 MW·h | 37.6% |
| ESBM1 | 0.72 | 27.17 MW | 15.29 MW | 5.23 MW·h | 57.0% |
| ESBM2 | 0.18 | 21.13 MW | 15.29 MW | 3.20 MW·h | 42.9% |
| ESBM3 | -0.33 | 12.80 MW | 10.24 MW | 1.69 MW·h | 30.8% |

Several observations of Table III are summarized below. First, in Table III, the discharge rate limit $n_+$ of ESBM1 increases to 27.17 MW, approximately 140% of the counterpart of the SBM and over 90% of that of the NBM. The resultant percentage utilization of wind power in the dispatch outcome is thus improved by 20%. This substantiates the conservativity of the SBM in [5] and the benefit of leveraging the ESBM.

Second, comparing the results related to the three ESBMs, it can be seen that the value of $\mu$ decreases as the line flow limits are increasingly tighter: ESBM1 with inactive line flow constraints has a $\mu$ of 0.72, but the $\mu$ of ESBM2 reduces to 0.18. The positive sign of $\mu$ suggests that the SBM in Table III is not sharp and covered by the RAF in these two cases. As for ESBM3, $\mu$ is -0.33, implying that the active line flow constraints have made the RAF shrink to $\mathbb{U}'$ that no longer covers the SBM that neglects the coupling constraints, so ESBM3 is generated by shrinking from the SBM. In this case, leveraging the SBM in the dispatch program is not conservative but risky, and leveraging the ESBM becomes even necessary to safeguard power system operations. Fig. 2 graphically summarizes the impact of the coupling constraints on the RAF for the three cases and the resultant ESBMs. Besides, the lowest percentage utilization of wind power regarding ESBM3 compared to the other cases highlights that accurately evaluating the aggregate flexibility of TCL arrays with possibly active coupling constraints is important to preventing from overestimating the power system's admissibility of wind power.

All in all, it is necessary to adopt our ESBM that is able to readily handle the coupling constraints for a more accurate evaluation of the aggregate flexibility of TCLs to improve and safeguard renewable-integrated power system operations.

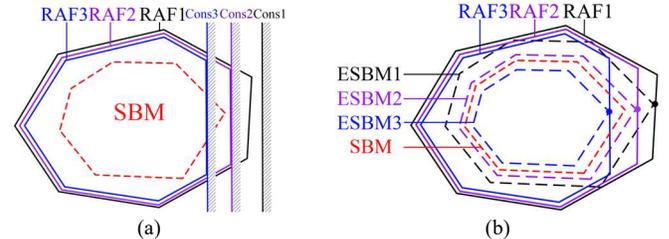

Fig. 2. Illustration of the coupling constraints' impact: (a) on the RAFs; (b) on the resultant ESBMs.